\documentclass{ws-ijmpd}
\usepackage{tabularx}
\usepackage{multirow}
\usepackage{url}
\usepackage[super,compress]{cite}
\begin{document}

\markboth{Xie et al.}
{Constraints on Cosmological Models with Gamma-Ray Bursts in Cosmology-Independent Way}

%
\catchline{}{}{}{}{}
%

\title{Constraints on Cosmological Models with Gamma-Ray Bursts in Cosmology-Independent Way}

\author{Hanbei Xie$\dag$, Xiaodong Nong$\dag$, Huifeng Wang, Bin Zhang, Zihao Li, and Nan Liang$^*$}

\address{Key Laboratory of Information and Computing Science Guizhou Province, Guizhou Normal University, Guiyang, Guizhou 550025, China\\
School of Cyber Science and Technology, Guizhou Normal University, Guiyang, Guizhou 550025, China\\
Joint Center for FAST Sciences Guizhou Normal University Node, Guiyang, Guizhou 550025, China\\
$\dag$Co-first author(s).
$^*$Corresponding author(s). E-mail(s): liangn@bnu.edu.cn}

\maketitle

\begin{history}
\received{Day Month Year}
\revised{Day Month Year}
\end{history}

\begin{abstract}
In this paper, we present a cosmology-independent method to constrain cosmological models from the latest 221 gamma-ray bursts (GRBs) sample, including 49 GRBs from Fermi catalog with the Amati relation  (the $E_{\rm p}$-${E}_{\rm iso}$ correlation), which are calibrated by using a Gaussian process from the Pantheon+ type Ia supernovae (SNe Ia) sample. 
With 182 GRBs at $0.8\le z\le8.2$ in the Hubble diagram and the latest observational Hubble data (OHD) by the Markov Chain Monte Carlo (MCMC) method, we obtained $\Omega_{\rm m}$ = $0.348^{+0.048}_{-0.066}$ and $h$ = $0.680^{+0.029}_{-0.029}$  for the flat $\Lambda$CDM model, and  $\Omega_{\rm m}$ = $0.318^{+0.067}_{-0.059}$, $h$ = $0.704^{+0.055}_{-0.068}$, $w$ = $-1.21^{+0.32}_{-0.67}$ for the flat $w$CDM model.
These results are consistent with those 
in which the coefficients of the Amati relation and the cosmological parameters fitted simultaneously.

\end{abstract}


\ccode{PACS numbers:}


\section{Introduction}	
Long Gamma-ray bursts (GRBs) are the most intense and energetic bursts of gamma rays from the cosmic space within a short period of time. Currently, the maximum observable redshift of GRBs is estimated to be around $z=9.4$ \cite{Cucchiara2011}, which is significantly greater than Type Ia supernovae (SNe Ia), with the maximum observable redshift $z\sim2.3$ \cite{Scolnic2018,Scolnic2022}.
Several empirical GRB luminosity relations, which are connections between measurable properties of the instantaneous gamma-ray emission and the luminosity or energy, e. g., the luminosity-variability relation\cite{Fenimore2000}, the luminosity-spectral lag relation\cite{Norris2000},  the Amati relation\cite{Amati2002},  the Yonetoku relation\cite{Yonetoku2004}, the Ghirlanda relation\cite{Ghirlanda2004a}, the Liang-Zhang relation\cite{Liang2005},
the Dainotti relation\cite{Dainotti2008}, and the Combo relation\cite{Izzo2015}, and the improved Amati relations from Gaussian Copula\cite{Liu2022a}, have been proposed to standardize GRBs \cite{Dai2004,Yonetoku2004,Ghirlanda2004b,Liang2005,Xu2005,Firmani2005,Firmani2006,Ghirlanda2006,Schaefer2007,Liang2008,Capozziello2008,Capozziello2009,LZ2008,Kodama2008,Tsutsui2009a,Liang2010,Liang2011,Gao2012}.
Among the luminosity relations of GRBs, the Amati relation \cite{Amati2002}, which connects the spectral peak energy and the isotropic equivalent radiated energy (the $E_{\rm p}$-${E}_{\rm iso}$ correlation), has been widely used to investigate cosmological parameters\cite{Amati2008,Capozziello2010,Wei2009,Wei2010,Liu2015,Wang2016,Demianski2017a,Demianski2021,Shirokov2020,Muccino2021,Lovyagin2022,Tang2022,Amati2019,Dirirsa2019,Khadka2020,Khadka2021,Montiel2021,AA2020,Galaxy2021,Luongo2021,Luongo2023,Muccino2023,Cao2022MN510,Liang2022,Liu2022,Jia2022,LZL2023,Kumar2023,Mu2023,Zhang2312,Shah2024,WLL2024,WL2024}.
For a general review, see Ref.~\refcite{Galaxy2021}.

However, the early studies had usually calibrated the luminosity relations of  GRBs  by assuming a FIDUCIAL cosmological model \cite{Dai2004,Schaefer2007}. 
Therefore, using these model-dependent GRB data to constrain cosmological models leads to the circularity problem \cite{Ghirlanda2006}.
Liang \emph{et al.}\cite{Liang2008} proposed  a model-independent method  to calibrate the luminosity relations of GRBs with SNe Ia data by the interpolation method and construct the GRB Hubble diagram, which can be used to constrain cosmological models \cite{Capozziello2008,Capozziello2009,Wei2009,Wei2010,Liang2010,Liang2011,Gao2012,Wang2016,Liu2022}. The luminosity relations of GRBs can be calibrated with SNe Ia data by the similar methods \cite{LZ2008,Kodama2008,Capozziello2010,Gao2012,Liu2015,Izzo2015,Demianski2017a,Demianski2021,Shirokov2020,Muccino2021,Lovyagin2022,Tang2022,Liang2022,Liu2022}.

Furthermore, the observational Hubble data (OHD) obtained with the cosmic chronometers (CC) method, which related the evolution of differential ages of passive galaxies at different redshifts \cite{Jimenez2002,Jimenez2003},
have unique advantages to calibrate GRBs in a model-independent way.
Amati \emph{et al.} \cite{Amati2019} proposed an alternative method to calibrate GRB correlations by using the OHD through the B\'ezier parametric curve and built up a Hubble diagram consisting of 193 GRBs with the Amati relation 
Following this method \cite{Amati2019}, 
several works have  constrained  cosmological models with the Amati relation 
\cite{Montiel2021,AA2020,Galaxy2021,Luongo2021,Luongo2023,Muccino2023}.

On the other hand, the simultaneous fitting method, in which the coefficients of relations and the parameters of the cosmological model are constrained simultaneously, has been proposed to avoid the circularity problem  \cite{Amati2008}.
Khadka \emph{et al.} \cite{Khadka2021} compile a data set of 118 GRBs (the A118 sample \cite{Khadka2020}, including 25 Fermi GRB sample \cite{Dirirsa2019}) with the smallest intrinsic dispersion from the total 220 GRBs (the A220 sample) with the Amati relation to derive the correlation and cosmological model parameters simultaneously. 
Cao \emph{et al.} \cite{Cao2022MN510,Cao2022MN512,Cao2022MN516} used the Amati relation \cite{Amati2002} with the A220 and the A118 GRB samples in conjunction with the Dainotti relation\footnote{Compared to GRB relations of the prompt emission phase, the relations involving the X-ray afterglow plateau phase \cite{Dainotti2008,Cardone2009,Cardone2010} exist less variability in its features. Dainotti \emph{et al.} proposed the relation between the plateau luminosity and the end time of the plateau in X-ray afterglows (2D Dainotti relation\cite{Dainotti2008}), 
which has been used for cosmological constraint \cite{Dainotti2010,Dainotti2011a,Dainotti2011b,Dainotti2013a,Dainotti2013b,Dainotti2015a,Dainotti2015b,Dainotti2017a}. Furthermore, the GRB  Fundamental Plane relation (the 3D Dainotti relation) among the rest-frame time and X-ray luminosity at the end of the plateau emission and the peak prompt luminosity 
with small intrinsic scatter has been found \cite{Dainotti2016,Dainotti2017b,Dainotti2020a,Dainotti2021a,Srinivasaragavan2020}. Some similar 2D and 3D relations  with the plateau in the X-ray afterglows have also been found \cite{Hu2021,Wang2022,Li2023}.
Recently, the relationship in optical wavelengths between the optical rest-frame end time and the optical luminosity at the end of the plateau has been found \cite{Dainotti2020b}.
Very recently, the GRB relation  in radio plateau phase afterglows  has also been investigated \cite{Levine2021,Tian2023}.} 
to constrain cosmological model parameters by the simultaneous fitting method.
They showed that Platinum sample including 50 GRB data could be standardized with a cosmological-model-independent 3D Dainotti relation\cite{Cao2022MN512}, and  the 3D Dainotti is strongly favored over the 2D one with different GRB data compilation \cite{Cao2022MN516}. 
Dainotti \emph{et al.} \cite{Dainotti2022MN514,Dainotti2022ApJS261} used optical and X-ray GRB fundamental planes as cosmological distance indicators.
Dainotti \emph{et al.} \cite{Dainotti2023MN518} corrected the 3D relation by considering the selection and evolutionary effects with a reliable statistical method to obtain a lower central value for the intrinsic scatter.
The 3D Dainotti relation has also been used  with a binned analysis with GRBs, SNe Ia, and baryonic acoustic oscillations (BAOs)  \cite{Dainotti2022PASJ74}; and joint constraints combined GRBs with quasars, SNe Ia, and BAOs \cite{Dainotti2023ApJ951,Bargiacchi2023MN521};
as well as a robust cosmographic technique \cite{Bargiacchi2023MN525}.

Up to now, whether the luminosity relations of GRB  are redshift dependent or not is still under debate.
The possible evolutionary effects in GRB relations have been discussed in many works \cite{Basilakos2008,Lin2016,Wang2017,Demianski2017a,Demianski2021,Dai2021,Tang2021,Dainotti2022MN514,Dainotti2023MN518}.
Regarding the luminosity function and density rate and cosmological evolution of the formation rate of GRBs, the luminosity relations of GRB could be evolved with redshift\cite{Petrosian2015,Lloyd2019,Tstetova2017,Yu2017,Dainotti2021ApJ914}.
With the A220 sample, Khadka \emph{et al.} \cite{Khadka2021} found that the Amati relation is independent
of redshift within the error bars; Liu \emph{et al.} \cite{Liu2022} proposed the improved Amati relation by accounting for evolutionary effects via copula, and  
found that a redshift evolutionary correlation is favored slightly;
Kumar \emph{et al.} \cite{Kumar2023}  divided the GRB data into five distinct redshift bins to calibrate the Amati relation, and found that GRBs do seem to evolve with cosmological redshift.

Recently,
Liang \emph{et al.} \cite{Liang2022} calibrated the Amati relation with the A219 sample and the A118 sample by using a Gaussian process from the Pantheon samples with 1048 SNe Ia data points \cite{Scolnic2018} and constrained cosmological models in flat space with GRBs at high redshift and 31 OHD via the Markov Chain Monte Carlo (MCMC) numerical method. 
Li, Zhang \& Liang \cite{LZL2023} calibrated GRB from the latest 32 OHD via the Gaussian process to construct the GRB Hubble diagram with the A118 data set and constrained Dark Energy models with GRBs at high redshift and SNe Ia in a flat space by the MCMC method.
Mu \emph{et al.} \cite{Mu2023} reconstruct cosmography parameters up to fifth order with the Amati relation of the A219 sample \cite{Liang2022} calibrated from Pantheon+ samples \cite{Scolnic2022}, which contains 1701 SNe light curves of 1550 spectroscopically confirmed SNe Ia at redshift $z<2.26$. 

More recently, Jia \emph{et al.} \cite{Jia2022}  found no statistically significant evidence for the redshift evolution  with the Amati relation from the analysis of data in different redshift intervals with a long GRB sample, which is the latest GRB sample so far, including 221 long GRBs  (the J221 sample) with redshifts from 0.03 to 8.20, including 49 GRBs updated from Fermi catalog (from 2013 to March 2021). 

In this paper, we utilize the latest 221 GRB data compiled in Ref.~\refcite{Jia2022} and the Pantheon+ sample \cite{Scolnic2022} to calibrate the Amati relation by Gaussian process at low redshift and obtain the  Hubble diagram of GRBs. 
We constrain cosmological models  with the GRBs at high redshift  and the latest 32 OHD data \cite{LZL2023}  by the MCMC method. Finally, we also use the simultaneous fitting method for constraints on cosmological models.

\section{CALIBRATION OF AMATI RELATION AND GRB Hubble DIAGRAM}
The Amati relation \cite{Amati2002}, which connects the spectral peak energy and the isotropic equivalent radiated energy (the $E_{\rm p}$-${E}_{\rm iso}$ correlation) of GRBs, can be expressed as
\begin{equation}y= a + bx,\end{equation}
where $y={\rm log}_{10}\frac{E_{{\rm is o}}}{{\rm 1erg}}$, $x={\rm log}_{10}\frac{E_{{\rm p}}}{{\rm 300keV}}$,
and $a$ and $b$ are free coefficients, 
$E_{{\rm iso}}$ and $E_{\rm p}$ can be calculated by
\begin{equation}E_{{\rm iso}} = 4\pi d^2_L(z)S_{{\rm bolo}}(1+z)^{-1},\quad E_{\rm p} = E^{{\rm obs}}_{{\rm p}}(1+z), \end{equation}
where $E^{{\rm obs}}_{{\rm p}}$ and  $S_{{\rm bolo}}$ are the GRB spectral peak energy and bolometric fluence.
It should be noted that the values of $E_{{\rm iso}}$ from Tab. 1 in Ref.~\refcite{Jia2022} are related to luminosity distance $d_L$, which depends on cosmological models.  The luminosity distance can be calculated by $ d_{\mathrm{L}}(z)= \frac{c(1+z)}{H_{0}}\int_{0}^{z} \frac{\mathrm{d} z^{\prime}}{\sqrt{\Omega_{m}(1+z^{\prime})^{3}+\Omega_{\mathrm{\Lambda}}}}$, where $\Omega_{\rm m}$ represents the parameter for non-relativistic matter density, $\Omega_{\Lambda}$ represents the cosmological constant density, and $H_0$ represents the Hubble constant.

\subsection{Reconstruction through Gaussian process from SNe Ia data}
Jia \emph{et al.} \cite{Jia2022}  used the standard cosmological parameters from Plank Collaboration \cite{Plank2020} ($\Omega_{\rm m}$ = 0.315, $\Omega_{\Lambda}$ = 0.685, and $H_0$ = 67.4 km $s^{-1}$ ${{\rm Mpc}}^{-1}$) without any error  to obtain the values of $E_{{\rm iso}}$.
In this work, we use Gaussian process method 
to obtain the values of $E_{{\rm iso}}$ without assuming a specific cosmological model.
Gaussian process, which is a fully Bayesian method for smoothing data to effectively reduce the errors of reconstructed results \cite{Seikel2012a,Seikel2012b}, has been widely applied to the field of cosmology \cite{Lin2018,Li2021,Sun2021,Wang2022,Liang2022,LZL2023,Benisty2023}. In order to obtain model-independent $E_{{\rm iso}}$, we use a Gaussian process to reconstruct the values of the luminosity distance ($d_L$) of GRBs  from SNe Ia data. In the Gaussian process, the function values $f(z)$  are correlated by a covariance function $k(z,\tilde z)$ to characterize the connection between the function values at different reconstructed points.
We adopt a squared exponential covariance function with the property of infinite differentiability suitable for reconstructing the shape of the function, which is given by  proposed by Seikel \emph{et al.} \cite{Seikel2012a}, $k(z,\tilde{z})=\sigma_f^2\exp\left[-\frac{(z-\tilde{z})^2}{2l^2}\right]$, where  $\sigma_f$ and $l$ are the hyperparameters need to optimize the values.

\begin{figure}
\centering
\includegraphics[width=180px]{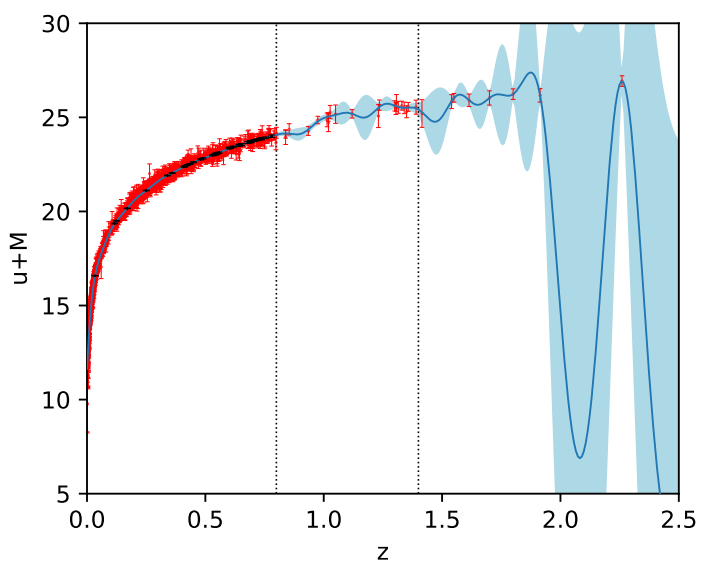}
\caption{The apparent magnitudes reconstructed through the Gaussian process from SNe Ia data at $z \le 2.26$. The blue curves present the reconstructed function with the 1$\sigma$ uncertainty from the SNe Ia data (red dots). The apparent magnitudes of GRBs at $z<0.8$ (black dots) are reconstructed from SNe Ia. The dashed line denotes $z = 0.8$ and $z = 1.4$. } \label{fig/1.png}
\end{figure}

In order to reconstruct GRB data  through Gaussian process from SNe Ia, we use the public Python package GaPP\footnote{\url{https://github.com/astrobengaly/GaPP}} with  the J221 GRB data\footnote{We revisited the J221 sample \cite{Jia2022}, which consists of 49 GRBs from Fermi catalog, 33 GRBs from Ref.~\refcite{Amati2019} and 139 GRBs from Ref.~\refcite{Wang2016}.
} and the Pantheon+ sample \cite{Scolnic2022} comprising 1701 light curves of 1550 unique spectroscopically confirmed SNe Ia.\footnote{The Pantheon+ sample \cite{Scolnic2022} do not use SNe from SNLS at $z > 0.8$ due to sensitivity to the $U$ band in model training, so the Pantheon+ statistics between $0.8 < z < 1.0$ are lower than that of Pantheon \cite{Scolnic2018} and the Joint Light-curve Analysis (JLA \cite{Betoule2014}).}
The distance modulus $\mu$ is related to the luminosity distance $d_L$: $\mu= m - M = 5\log \frac{d_L}{\textrm{Mpc}} + 25$, where $m$ and $M$ correspond to the apparent magnitude and absolute magnitude, respectively. The reconstructed apparent magnitudes from Pantheon+ sample are shown in Fig. \ref{fig/1.png}. We find that the SNe Ia data points are sparse at $0.8\leq z\leq2.26$, the reconstruction function exhibits strange oscillations and large uncertainty.
Therefore, we use the reconstruction function at $z<0.8$ with 39 GRBs in J221 sample to calibrate the Amati relation.
In order to compare with the previous analyses \cite{Liu2022a,Liang2022,LZL2023}, we also used a subsample of SNe Ia with a redshift cutoff at $z=1.4$ to calibrate the Amati relation with 90 GRBs  at $z<1.4$ from the J221 sample. 

\subsection{Calibration of Amati Relation at low redshift}
We perform MCMC method\cite{ForemanMackey2013} by employing the Python package \texttt{emcee}\footnote{\url{https://emcee.readthedocs.io/en/stable/}} for fitting the Amati relation with GRBs in J221 sample at low redshift.
Two likelihood functions by D'Agostini\cite{D'Agostini2005} and Reichart \cite{Reichart2001} 
are used to fit the parameters of Amati relation ($a$ and $b$). The likelihood function proposed by Ref.~\refcite{D'Agostini2005} is written as
\begin{eqnarray}\label{Lc}
    \mathcal{L}_{\rm D}\propto\prod_{i=1}^{N_1} \frac{1}{\sigma^2}
    \times\exp\left[-\frac{[y_i-y(x_i,z_i; a, b, M)]^2}{2\sigma^2}\right].
\end{eqnarray}
Here $\sigma=\sqrt{\sigma_{\rm int}^2+\sigma_{y,i}^2+b^2\sigma_{x,i}^2}$,  $\sigma_{\rm int}$ is the intrinsic scatter of GRBs, $\sigma_y=\frac{1}{\rm ln10}\frac{\sigma_{E_{\rm iso}}}{E_{\rm iso}},\quad \sigma_x=\frac{1}{\rm ln10}\frac{\sigma_{E_{\rm p}}}{E_{\rm p}}$, $\sigma_{E_{\rm p}}$ is the error magnitude of $E_{\rm p}$, and $\sigma_{E_{\rm iso}}=4\pi d^2_L\sigma_{S_{\rm bolo}}(1+z)^{-1}$ is the error magnitude of $E_{\rm iso}$, where $\sigma_{S_{\rm bolo}}$ is the error magnitude of $S_{\rm bolo}$. It should be noted that the use of the Ref.~\refcite{D'Agostini2005} likelihood  may introduce a subjective bias on the choice of the independent variable in the analysis. The likelihood function proposed by Ref.~\refcite{Reichart2001} has the advantage of not requiring the arbitrary choice of an independent variable among $E_{p}$ and $E_{{\rm iso}}$, which has been used to get rid of this bias \cite{Amati2013,LZL2023}.
The Ref.~\refcite{Reichart2001} likelihood function can be written as \cite{Lin2016,LZL2023}
\begin{eqnarray}\label{eqnarray6}
    \mathcal{L}_{\rm R}\propto\prod_{i=1}^{N_1} \frac{\sqrt{1+b^2}}{\sigma}
    \times\exp\left[-\frac{[y_i-y(x_i,z_i; a, b, M)]^2}{2\sigma^2}\right]
\end{eqnarray}
The intrinsic scatter can be calculated by $\sigma_{\rm int}=\sqrt{\sigma_{y,\rm int}^2 + b^2\sigma_{x,\rm int}^2}$, in which $\sigma_{x,\rm int}$ and $\sigma_{y,\rm int}$ are the intrinsic scatter along the $x$-axis and $y$-axis.

The calibrated results (the intercept $a$, the slope $b$, the intrinsic scatter $\sigma_{\rm int}$, and the absolute magnitude $M$ of SNe Ia) in the the J221 sample with redshift  $z < 0.8$ (39 GRBs) and $z < 1.4$ (90 GRBs) are summarized in Tab. \ref{ta1}.
We find that the fitting results of the intercept ($a$) with the two likelihood function methods \cite{D'Agostini2005,Reichart2001} are consistent in 1$\sigma$ uncertainty; however,
there is a significant difference in the slope parameter $b$ with the two likelihood function methods \cite{D'Agostini2005,Reichart2001}. As pointed out in Ref.~\refcite{LZL2023}, this discrepancy arises because the likelihood employed by Ref.~\refcite{D'Agostini2005} may introduce subjective biases. 

\begin{table}
\tbl{Calibration results (the intercept $a$, the slope $b$, the intrinsic scatter $\sigma_{\rm int}$ and the absolute magnitude $M$) of the Amati relation in the J221 GRB sample at $z < 0.8$ and $z < 1.4$ by the likelihood method Reichart 2001\cite{Reichart2001} and the likelihood method D'Agostini 2005\cite{D'Agostini2005}.}
{\begin{tabular}{@{}cccccc@{}} \toprule
Methods & data sets &$a$& $b$& $\sigma_{{\rm int}}$& $M$ \\ \hline
\multirow{2}{*}{ D'Agostini 2005 } &39GRBs ($z < 0.8$) & $52.75^{+0.58}_{-0.58}$ & $1.50^{+0.13}_{-0.13}$ & $0.431$ & $-19.50^{+0.14}_{-0.14}$\\
\multirow{3}{*}{} & 90GRBs($z < 1.4$) & $52.83^{+0.58}_{-0.58}$ & $1.59^{+0.10}_{-0.10}$ & $0.433$ & $-19.50^{+0.14}_{-0.14}$\\
\hline
\multirow{2}{*}{ Reichart 2001} &39GRBs ($z < 0.8$) & $52.80^{+0.47}_{-0.87}$ & $1.808^{+0.094}_{-0.120}$ & $0.413$ & $-19.40^{+0.14}_{-0.14}$\\
\multirow{3}{*}{} & 90GRBs($z < 1.4$)  & $52.87^{+0.58}_{-0.58}$ & $2.026^{+0.083}_{-0.093}$ & $0.423$ & $-19.50^{+0.14}_{-0.14}$\\
\botrule
\end{tabular} \label{ta1}}
\end{table}

\begin{figure}
      \includegraphics[width=175px]{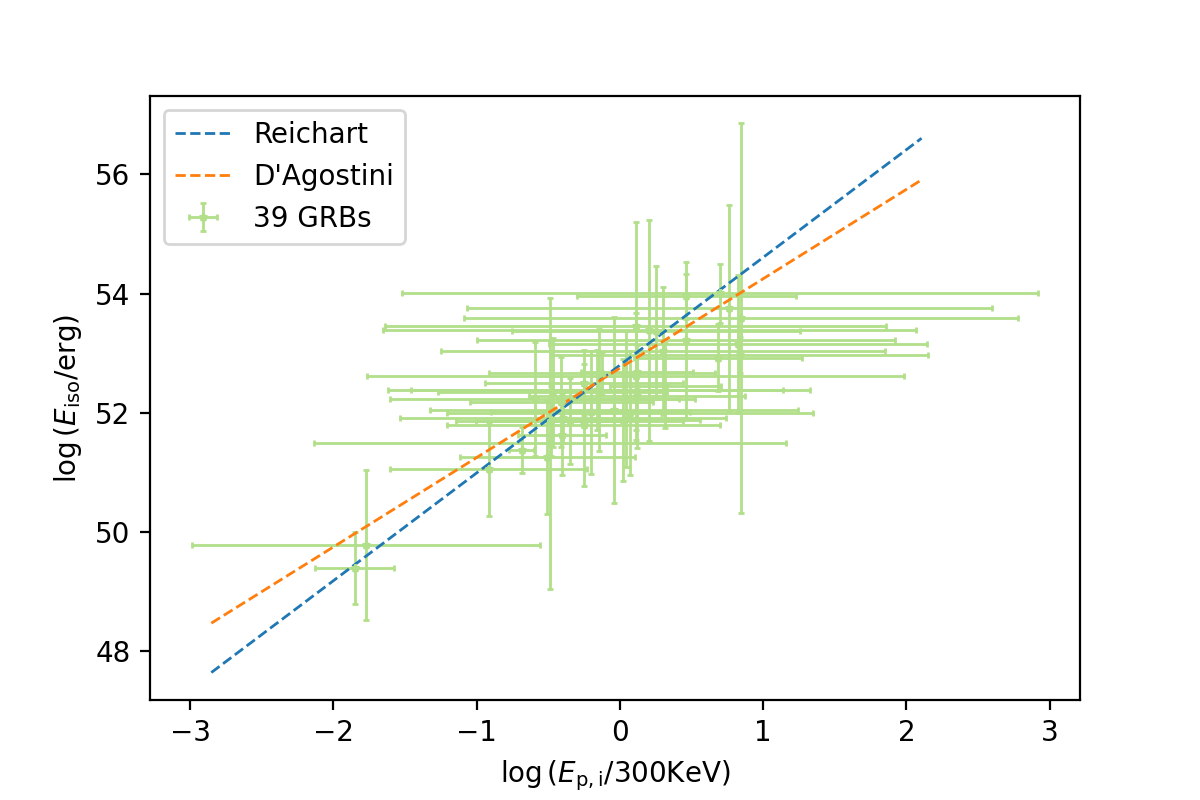}
      \includegraphics[width=175px]{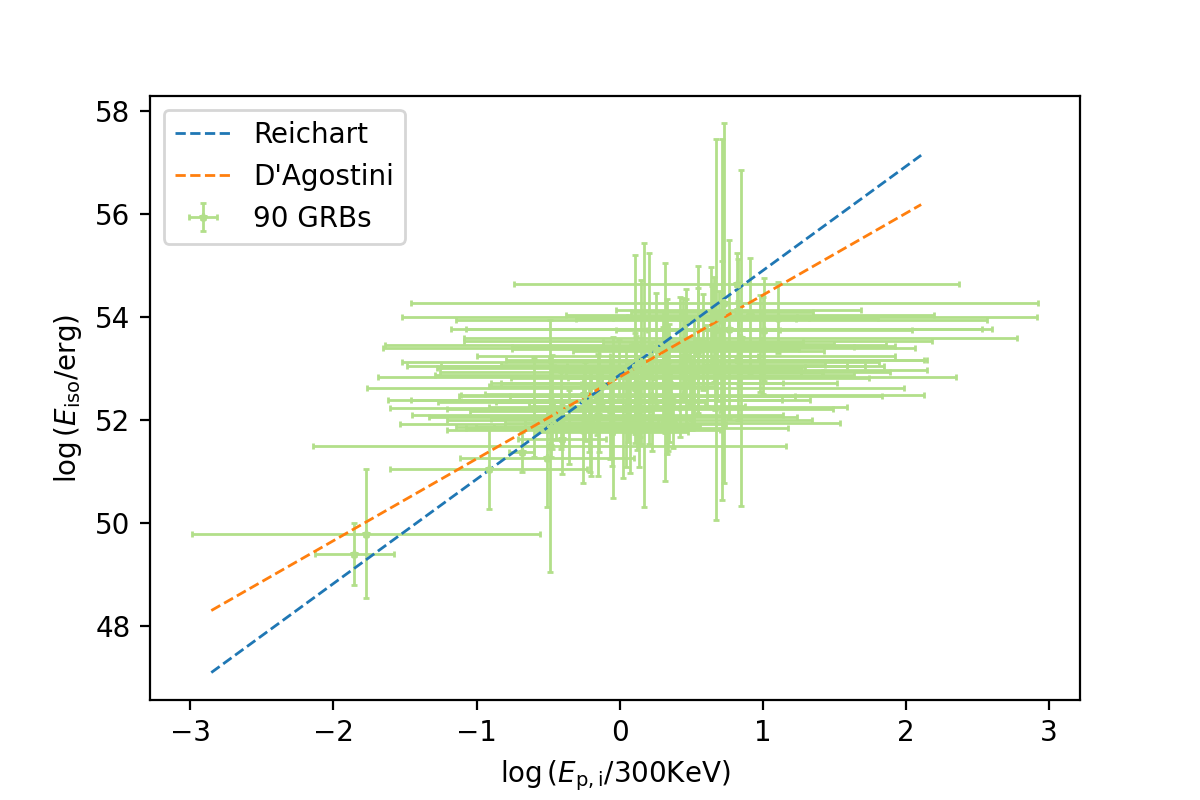}
      \caption{The best-fit calibration of Amati relation by the D'Agostini method (brown dashed line) and the Reichart method (blue dashed line) with redshifts $z < 0.8$ (39 GRBs, left panel) and $z < 1.4$ (90 GRBs, right panel). By the Reichart method, the Spearman$'$s rank coefficient of the correlation is $\rho = 0.85$, and the P-value = $2.88\times10^{-26}$ for the sample at $z < 0.8$; and $\rho = 0.87$, P-value = $1.12\times10^{-12}$ for the sample at $z < 1.4$.}
      \label{fig:amati}
    \end{figure}

We plot the Amati relation by D'Agostini method and Reichart method with redshifts $z < 0.8$ (39 GRBs) and $z < 1.4$ (90 GRBs) in Fig.\ref{fig:amati}.
We find significant scatter and large uncertainty in the $E_p$ measurements and $E_{\rm iso}$ estimations for several of GRB data at low redshift.
Obviously, 
the uncertainties of 221 GRBs  in Ref.~\refcite{Jia2022} which calibrated by recent results from
Planck Collaboration without any error can reduce scatter and uncertainties.
However, our results preserves the  original variability of GRB data without assuming a specific cosmological model, which can lead to larger intrinsic scatter.
Other samples from cosmological independent calibrations appear to show less scatter and lower uncertainties. For instance,
Ref.~\refcite{Amati2019} used a B\'ezier parametric curve for calibration with 193 GRBs from 31 OHD, which resulting in less scatter and smaller uncertainties in $E_{\rm iso}$.
The possible reason may due to the GP method from SNe Ia can lead to larger intrinsic scatter compared to the B\'ezier method from Hubble parameter data in reconstructions to calculate GRB luminosity distance and $E_{\rm iso}$.

We find that our results of slope by Reichart method ($b=1.808^{+0.094}_{-0.120}$ at $z<0.8$ and $b=2.026^{+0.083}_{-0.093}$ at $z<1.4$) are close to the typical value ($E_{\rm iso} \propto E_{\rm p}^2$) found in recent Fermi data \cite{WL2024} and previous 193 GRBs\cite{Amati2019}. The large uncertainty and the information coming from the huge scatter of GRBs do not seem to propagate to the slope determination. \footnote{Titarchuk \emph{et al.}\cite{Titarchuk2012} have given a physical interpretation of the Amati relation with a model for the spectral formation of GRB prompt emission with two upscattering processes (the GRBCOMP model). If the timescale of the GRB prompt emission and its shape for any burst is more or less the same, then $E_{\rm p} \propto E_{\rm iso}^{1/2}$ is seen precisely. The observable scattering of the Amati relation can be caused by the spread of the parameter values that characterize the outflow evolution of each GRB and the different temperature of the soft photons related to the emission of the star itself. Frontera \emph{et al.}\cite{Frontera2012}  have concluded that the Yonetoku relation (the spectral peak energy $E_{\rm p}$ - the isotropic bolometric luminosity $L_{\rm iso}$) is intrinsic to the emission process and their results strongly support the reality of the Amati and Yonetoku relations both derived using time-averaged spectra. Frontera \emph{et al.}\cite{Frontera2013} have used the same GRB data \cite{Frontera2012} to test different physical models and confirmed the physical interpretation of the Yonetoku relation given by GRBCOMP.}
We also find that the fitting results of the intercept ($a$) with the two likelihood function methods \cite{D'Agostini2005,Reichart2001} are consistent in 1$\sigma$ uncertainty; however,
there is a significant difference in the slope parameter $b$ with the two likelihood function methods \cite{D'Agostini2005,Reichart2001}. As pointed out in Ref.~\refcite{LZL2023}, this discrepancy arises because the likelihood employed by Ref.~\refcite{D'Agostini2005} may introduce subjective biases. 
To avoid any bias in the selection of independent variables, we utilize the calibration results obtained through the likelihood method proposed by Ref.~\refcite{Reichart2001} to construct the GRB Hubble diagrams.

\subsection{GRB Hubble Diagram}
\begin{figure}
\centering
\includegraphics[width=180px,clip]{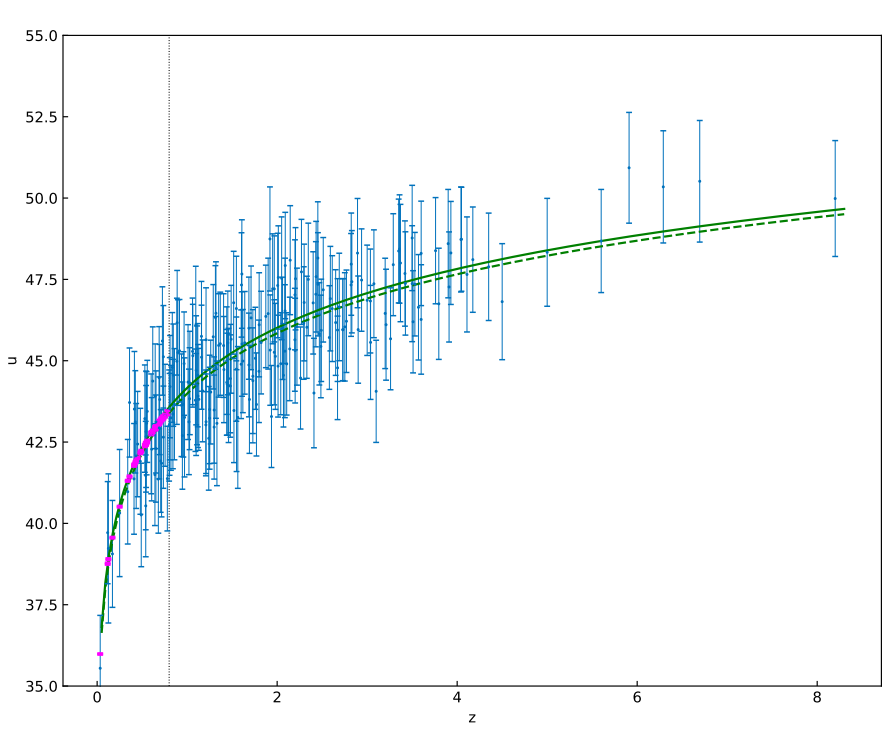}
\caption{GRB Hubble diagram with the J221 data set. GRBs at $z < 0.8$ were obtained by a Gaussian process from the SNe Ia data (purple points), while GRBs with $z \geq 0.8$ (blue points) were obtained by the Amati relation and calibrated with J221 at $z < 0.8$. The solid green curve presents the best-fit values from Plank CMB data at high-redshift: $H_0$ = 67.36 km $s^{-1}$ ${{\rm Mpc}}^{-1}$, $\Omega_m$ = 0.315 \cite{Plank2020}, and the green long dotted curve presents the best-fit constraints on the $\Lambda$CDM model from the Pantheon+: $H_0$ = 73.4 km $s^{-1}$, ${{\rm Mpc}}^{-1}$, $\Omega_m$ = 0.306 \cite{Scolnic2022,Brout2022,Riess2022}. The black dotted line denotes $z = 0.8$.} \label{2.png}
\end{figure}

To avoid any bias in the selection of independent variables, we utilize the calibration results obtained through the likelihood method proposed by Ref.~\refcite{Reichart2001} to construct the GRB Hubble diagrams.
Extrapolating the results from the low-redshift GRBs ($z < 0.8$, $z < 1.4$) from J221 sample to the high-redshift ones, we obtain the energy ($E_{iso}$) of each burst at high redshift ($z \geq 1.4$). Therefore, the luminosity distance ($d_L$) can be derived. The Hubble diagram with J221 sample \textbf{is} plotted in Fig.\ref{2.png}.

The uncertainty of GRB distance modulus with the Amati relation is
\begin{equation}\sigma^2_\mu=\bigg(\frac{5}{2}\sigma_{{\rm log}\frac{E_{\rm iso}}{\rm 1erg}}\bigg)^2+ \bigg (\frac{5}{\rm 2ln10}\frac{\sigma_{S_{\rm bolo}}}{S_{\rm bolo}} \bigg)^2 \ ,\end{equation}
where
\begin{equation}\sigma^2_{{\rm log}\frac{E_{\rm iso}}{\rm 1erg}}=\sigma^2_{\rm int}+ \bigg (\frac{b}{\rm ln10}\frac{\sigma_{E_{\rm p}}}{E_{\rm p}} \bigg )^2+\sum \bigg (\frac{\partial_{y}(x;\theta_c)}{\partial \theta_i} \bigg)^2C_{ii}\, .
\end{equation}
Here $\theta_c$ = \{$\sigma_{{\rm int}}$, $a, b$\}, and $C_{ii}$ means the diagonal element of the covariance matrix of these fitting coefficients.

\section{CONSTRAINTS ON COSMOLOGICAL MODELS}
The cosmological parameters can be fitted by minimizing the $\chi^2$ statistic.\begin{equation}\chi^2_{\rm GRB} = \sum^{N_1}_{i=1} \left[\frac{\mu_{\rm obs}(z_i)-\mu_{\rm th}(z_i;p,H_0)}{\sigma_{\mu_i}}\right]^2.
\end{equation}
Here, $N_1$ = 182 or 131 represents the number of high-redshift GRBs with $z \le 0.8$ or $z \le 1.4$, respectively, in the J221 sample, $\mu_{{\rm obs}}$ is the observational value of distance modulus and its error $\sigma_{\mu_i}$. $\mu_{{\rm th}}$ is the theoretical value of distance modulus calculated from the cosmological model,  $H_0$ is the Hubble constant,  $p$ represents the cosmological parameters.
For the recent CMB results for Planck alone: $\Omega_k = -0.0096\pm0.0061$ \cite{Plank2020}, we assume a flat space in this work. Considering the $\Lambda$CDM model, $w$CDM model, and the well-known  Chevallier-Polarski-Linder (CPL) model \cite{CP2001,L2003} with a DE component that depends on redshift, 

the theoretical value of the luminosity distance can be obtained by
\begin{equation}d_{L;{\rm th}}=\frac{{c}(1+z)}{H_{\rm 0}}\int^z_0\frac{dz}{\Omega_{\rm m}(1+z)^3+\Omega_{\rm DE}X(z)} . \end{equation}
Here $c$ is the speed of light, and $\Omega_{\rm m}$ and $\Omega_{DE}$ are the present dimensionless density parameters of matter and dark energy, respectively, which satisfy $\Omega_{\rm m}$ + $\Omega_{DE}$ = 1. $X(z)$ is determined by the choice of the specific DE model, it can be calculated by
\begin{eqnarray}\label{models}
	X(z)=\begin{cases}
		1, & \rm{\Lambda CDM} \\
		(1+z)^{3(1+w_0)},& w\rm{CDM} \\
		(1+z)^{3(1+w_0+w_a)}e^{-\frac{3w_az}{1+z}}. & \rm{CPL} \\
\end{cases}\end{eqnarray}

\begin{figure*}
\centering
\includegraphics[width=130px]{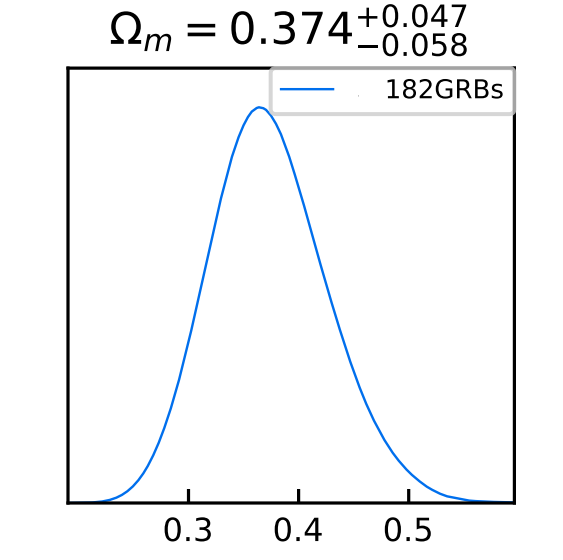}
\includegraphics[width=130px]{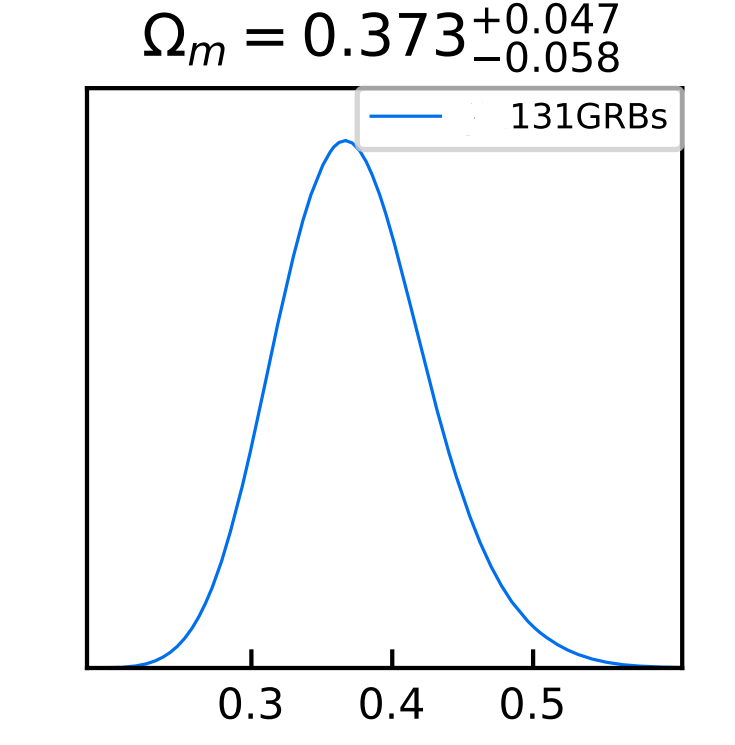}
\caption{Constraints on $\Omega_{\rm m}$ in the $\Lambda$CDM model at redshift $z \geq 0.8$ and $z \geq 1.4$, with 182 GRBs (left panel) and 131 GRBs (right panel) from the J221 GRBs dataset, respectively. $H_0$ is set to be $70\ {\rm km}\ {\rm s}^{-1}{\rm Mpc}^{-1}$ for the cases only with GRBs.} \label{constrain1}
\end{figure*}

\begin{figure*}
\centering
\includegraphics[width=170px]{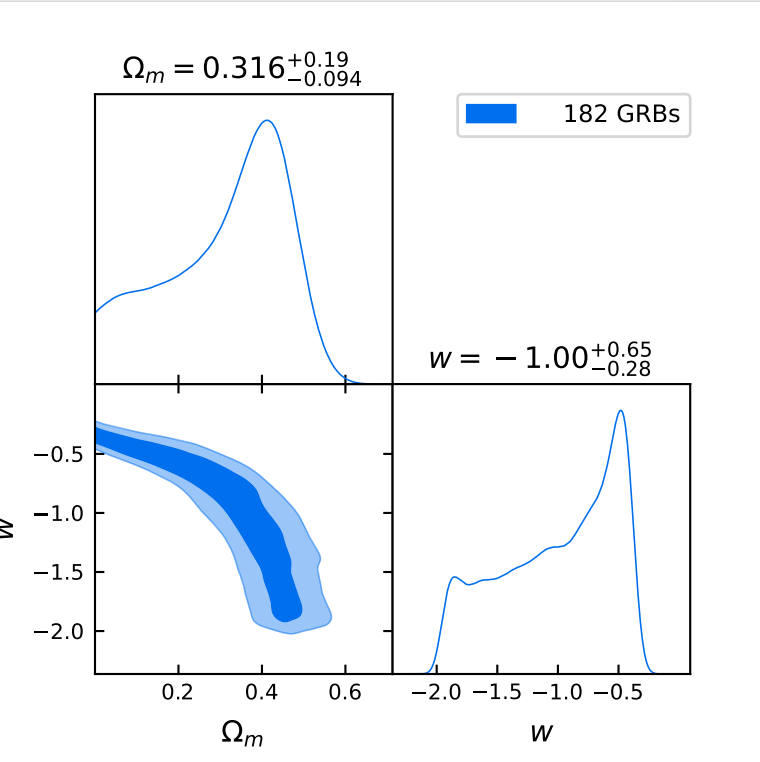}
\includegraphics[width=170px]{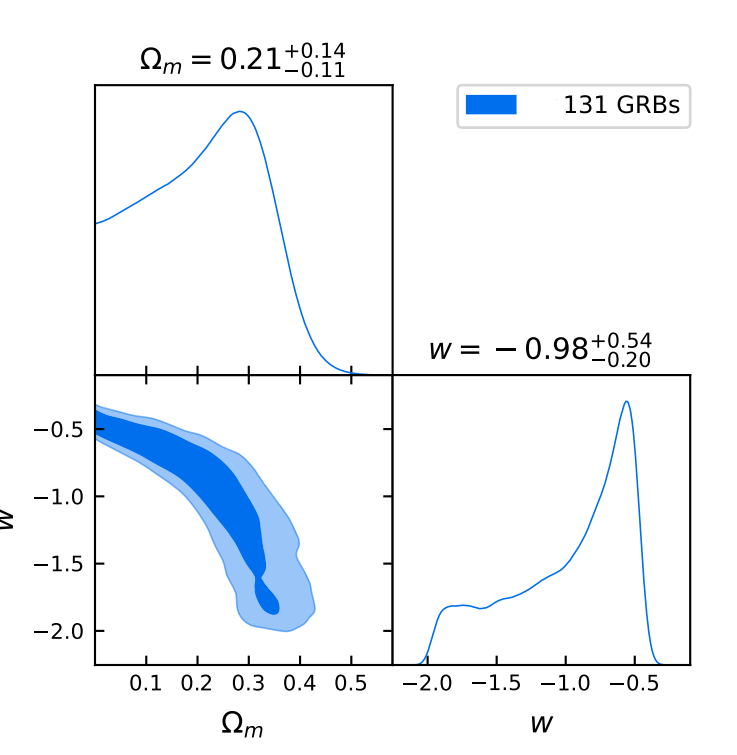}
\caption{Constraints on $\Omega_{\rm m}$ and $w$ in the $w$CDM model at redshift $z \geq 0.8$ and $z \geq 1.4$, with 182 GRBs (left panel) and 131 GRBs (right panel) from the J221 GRBs dataset, respectively. $H_0$ is set to be $70\ {\rm km}\ {\rm s}^{-1}{\rm Mpc}^{-1}$ for the cases only with GRBs.} \label{constrain2}
\end{figure*}

\begin{figure*}
\centering
\includegraphics[width=170px]{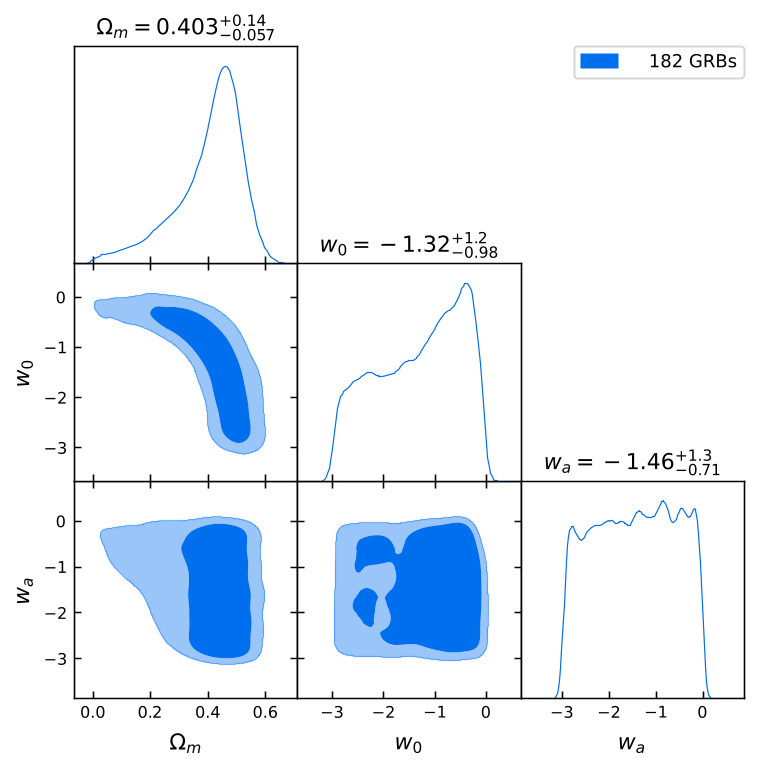}
\includegraphics[width=170px]{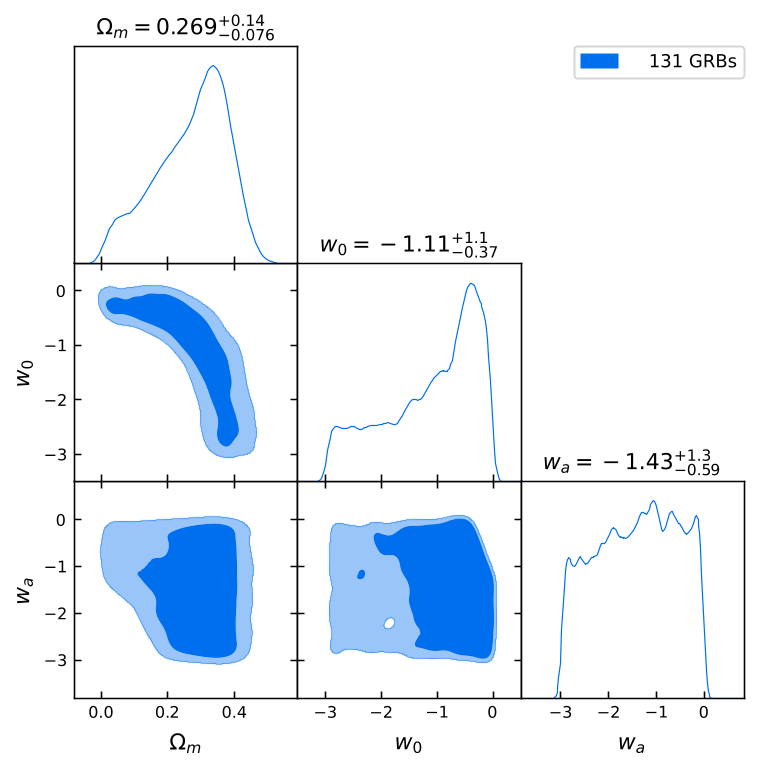}
\caption{Constraints on $\Omega_{\rm m}$, $w$ and $w_a$ in the CPL model at redshift $z \geq 0.8$ and $z \geq 1.4$, with 182 GRBs (left panel) and 131 GRBs (right panel) from the J221 GRBs dataset, respectively. $H_0$ is set to be $70\ {\rm km}\ {\rm s}^{-1}{\rm Mpc}^{-1}$ for the cases only with GRBs.} \label{constrain_CPL_fixed_h}
\end{figure*}

We employ the Python package \texttt{emcee} \cite{ForemanMackey2013} to constrain cosmological models with the GRB data at high-redshift. 
It should be noted that GRB data alone are unable to constrain $H_0$ because of the degeneracy between $H_0$ and the correlation intercept parameter; therefore $H_0$ is set to be $70\ {\rm km}\ {\rm s}^{-1}{\rm Mpc}^{-1}$ for GRB-only analyses in previous works \cite{Khadka2021, Cao2022MN512, Liang2022, LZL2023}.
The results of 182 and 131 GRBs in the J221 data set at $z \geq 0.8$ and $z \geq 1.4$ are shown in Fig. \ref{constrain1} ($\Lambda$CDM model), Fig. \ref{constrain2} ($w$CDM model). Constraint results with 1$\sigma$ confidence level are summarized in Tab. \ref{tab2} \footnote{Similar $\Omega_m$ value will provided when $H_0$ is treated as a free parameter in the fitting procedure\cite{Li2023}. For $H_0$ is coupled with the absolute magnitude $M$ of SNe Ia, $H_0$ is effectively constrained by the SNe Ia data if  $M$ is set as a concrete value in the low-redshift calibration\cite{Liu2022}.}.
With 182 GRBs at $0.8\le z\le8.2$ in the J221 sample, we obtained $\Omega_{\rm m}$ = $0.373^{+0.047}_{-0.058}$ ($\Lambda$CDM) and $\Omega_{\rm m}=0.316^{+0.190}_{-0.094}$, $w=-1.00^{+0.65}_{-0.28}$ ($w$CDM), which are consistent with Ref.~\refcite{Dainotti2023MN518} using the 3D GRB relation alone calibrated on SNe Ia ($\Omega_{\rm m}$ = $0.306\pm0.069$ for $\Lambda$CDM, fixing $h=70$; and $w$ = $-0.906\pm0.697$ for $w$CDM, fixing $\Omega_{\rm m}=0.3$, $h=70$). Our results are more stringent than previous results in Ref.~\refcite{Cao2022MN516} with the Platinum GRB and the LGRB95 sample for $\Lambda$CDM and $w$CDM model.
For the CPL model with GRBs at redshift $z \geq 0.8$ and $z \geq 1.4$, as shown in Fig. \ref{constrain_CPL_fixed_h}, we obtain 
$w_a=-1.46^{+1.30}_{-0.71}$, $w_a=-1.43^{+1.30}_{-0.59}$, which favors a possible DE evolution at the $1 \sigma$ confidence region. We also find that the values of $\Omega_{\rm m}$ obtained in this work for GRB-only analyses are much smaller than the early works\cite{Liang2022}; however, these results are consistent with ones from recent Fermi data\cite{WL2024}.

\begin{figure*}
\centering
\includegraphics[width=175px]{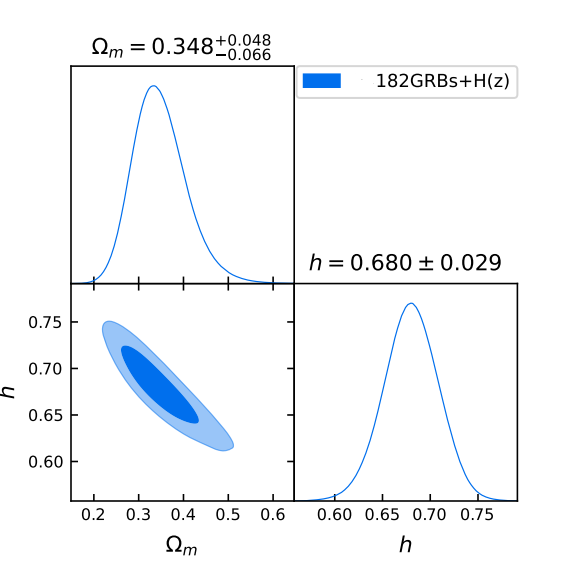}
\includegraphics[width=175px]{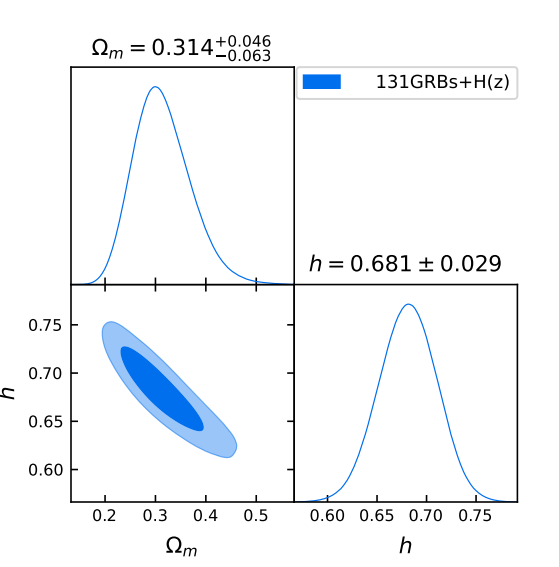}
\caption{Joint constraints on parameters of $\Omega_m$ and $h$ in the $\Lambda$CDM model at redshift $z \geq 0.8$ and $z \geq 1.4$, with 182 GRBs + 32 OHD (left panel) and 131 GRBs + 32 OHD (right panel), respectively from J221 GRBs dataset.}\label{constrain3} 
\end{figure*}

\begin{figure*}
\centering
\includegraphics[width=175px]{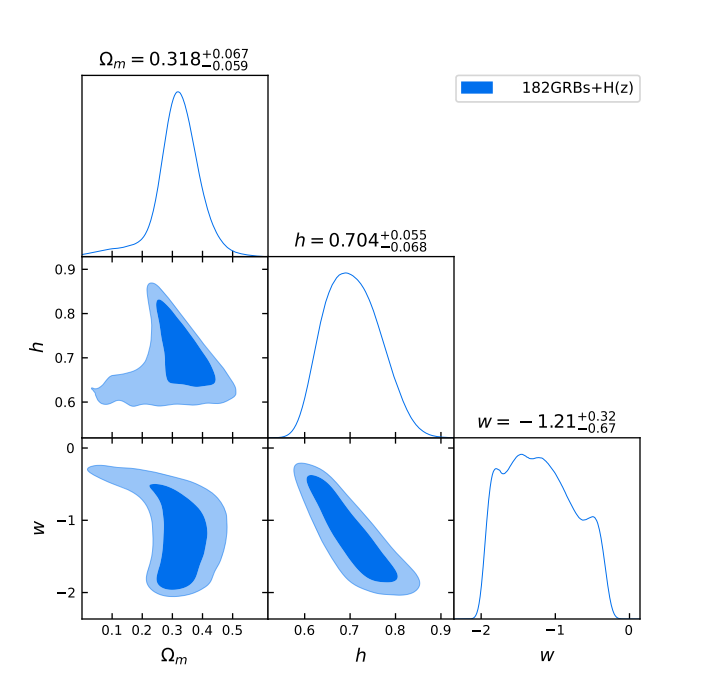}
\includegraphics[width=175px]{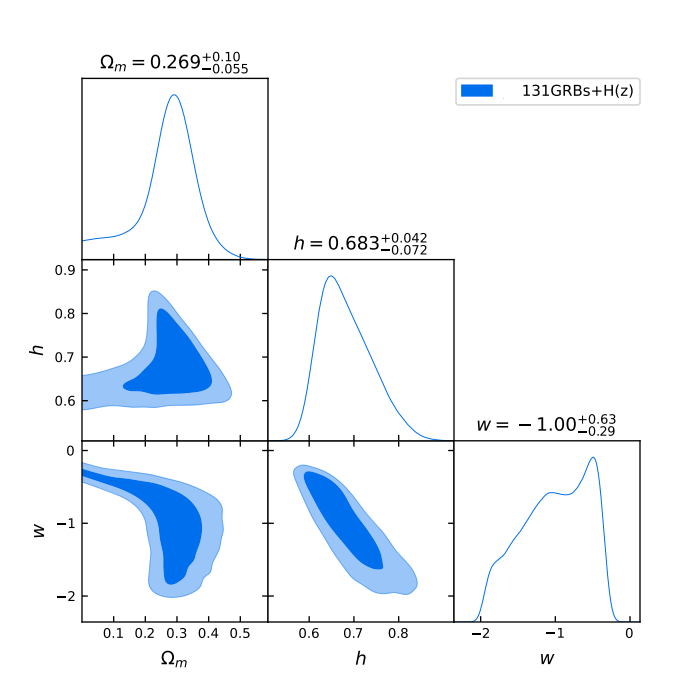}
\caption{Joint constraints on parameters of $\Omega_m$, $h$ and $w$ in the $w$CDM model at redshift $z \geq 0.8$ and $z \geq 1.4$, with 182 GRBs + 32 OHD (left panel) and 131 GRBs + 32 OHD (right panel), respectively from J221 GRBs dataset.}\label{constrain4}
\end{figure*}

It should be emphasized that GRBs cannot confidently constrain this DE model alone, as shown in Fig. \ref{constrain1}-\ref{constrain_CPL_fixed_h}.
Therefore, we also use the latest OHD in Ref.~\refcite{LZL2023} to constrain cosmological models, 
including the 31 Hubble parameter measurements at $0.07<z<1.965$ \cite{Stern2010,Moresco2012,Moresco2015,Moresco2016,Zhang2014,Ratsimbazafy2017} and a new point at $z=0.80$ proposed by Jiao \emph{et al.} \cite{Jiao2023} in a similar approach. 
It should be noted that Borghi \emph{et al.}\cite{Borghi2022} obtained  another new OHD at $z=0.75$. Considering these two measurements \cite{Jiao2023,Borghi2022} are
not fully independent and their covariance is not clear, we only use the point in Ref.~\refcite{Jiao2023},  which takes advantage of the $~1/\sqrt 2$ fraction of systematic uncertainty. One could either use the data in Ref.~\refcite{Borghi2022} alternatively  with other 31 OHD to investigate cosmology \cite{Kumar2023,Muccino2023,Lee2023,Favale2023}.
For the OHD data set, the $\chi^2$ has the form \cite{ZLYNLW2024}
\begin{equation}
\chi^2_{\text{OHD}} = \sum_{i=1}^{17} \left( \frac{H_{\text{obs}}(z_i) - H_{\text{th}}(z_i; p, H_0)}{\sigma_{H_i}} \right)^2 + \Delta \hat{H}^T C^{-1}_H \Delta \hat{H}
\end{equation}
where $\sigma_{H_i}$ represents the observed uncertainty of the 17 uncorrelated measurements; $\Delta \hat{H} = H_{\text{obs}}(z_i) - H_{\text{th}}(z_i; p, H_0)$ represents the difference 
of the 15 correlated measurements with the inverse of the covariance matrix $C^{-1}_H $. The full (non-diagonal) covariance matrix \cite{Moresco2020} for the correlated measurements is computed as\footnote{\url{https://gitlab.com/mmoresco/CCcovariance}}:
\begin{eqnarray}\label{eq:cov1}
	C_{H} = C_{\text{stat}} + C_{\text{sys}}.
\end{eqnarray}
Here, $C_{\text{stat}}$ incorporates statistical errors and is diagonal, and $C_{\text{sys}}$ encompasses systematic uncertainties related to the estimation of physical properties of galaxies, such as stellar metallicity and potential contamination by a young component.
The total $\chi^2$ of GRB and OHD data is
\begin{equation}
    \chi^2_{{\rm total}} = \chi^2_{{\rm GRB}} + \chi^2_{{\rm OHD}}.
\end{equation}
The constraint results of the high-redshift GRBs (182 GRBs at $z\geq 0.8$, and 131 GRBs at $z\geq 1.4$) from the J221 data set and 32 OHD, are plotted in Fig. \ref{constrain3} ($\Lambda$CDM model), Fig. \ref{constrain4} ($w$CDM model), and summarized in Tab. \ref{tab2} with the 1$\sigma$ confidence level.

With 182 GRBs at $0.8 \leq z \leq 8.2$ in the J221 sample and 32 OHD, we obtained $\Omega_{\rm m}=0.348^{+0.048}_{-0.066}$ and $h=0.680^{+0.029}_{-0.029}$ for the flat $\Lambda$CDM model, and $\Omega_{\rm m}$ = $0.318^{+0.067}_{-0.059}$, $h$ = $0.704^{+0.055}_{-0.068}$, $w$ = $-1.21^{+0.32}_{-0.67}$ for the flat $w$CDM model, which are consistent with the results using the 193 GRBs (Amati relation) and SNe Ia ($\Omega_{\rm m}=0.397\pm0.040$  for the $\Lambda$CDM model, and $\Omega_{\rm m}=0.34^{+0.13}_{-0.15}, w=-0.86^{+0.36}_{-0.38}$ for the $w$CDM model;  fixing $h=0.6774$) at the 2$\sigma$ level \cite{Amati2019}; and the results in Ref.~\refcite{Dainotti2022MN514} combining SNe Ia and GRBs with a 3D optical Dainotti correlation for a flat $\Lambda$CDM cosmology ($\Omega_{\rm m}=0.299\pm0.009)$.
With 131 GRBs at $1.4\le z\le8.2$ in the J221 sample and 32 OHD, we obtained $\Omega_{\rm m}$ = $0.314^{+0.046}_{-0.063}$ and $h$ = $0.681^{+0.029}_{-0.029}$ for the flat $\Lambda$CDM model, and  $\Omega_{\rm m}$ = $0.269^{+0.100}_{-0.055}$, $h$ = $0.683^{+0.042}_{-0.072}$, $w$ = $-1.00^{+0.63}_{-0.29}$ for the flat $w$CDM model at the 1$\sigma$ confidence level, which are consistent with our previous analyses with 98 GRBs at $1.4<z\leq8.2$ in the A118 sample and OHD ($\Omega_{\rm m}$=$0.346^{+0.048}_{-0.069}$, $h$=$0.677^{+0.029}_{-0.029}$ for the flat $\Lambda$CDM model, and $\Omega_{\rm m}$=$0.314^{+0.072}_{-0.055}$, $h$=$0.705^{+0.055}_{-0.069}$, $w$=$-1.23^{+0.33}_{-0.64}$ for the flat $w$CDM model)\cite{Liang2022}.

\begin{figure*}
\centering
\includegraphics[width=175px]{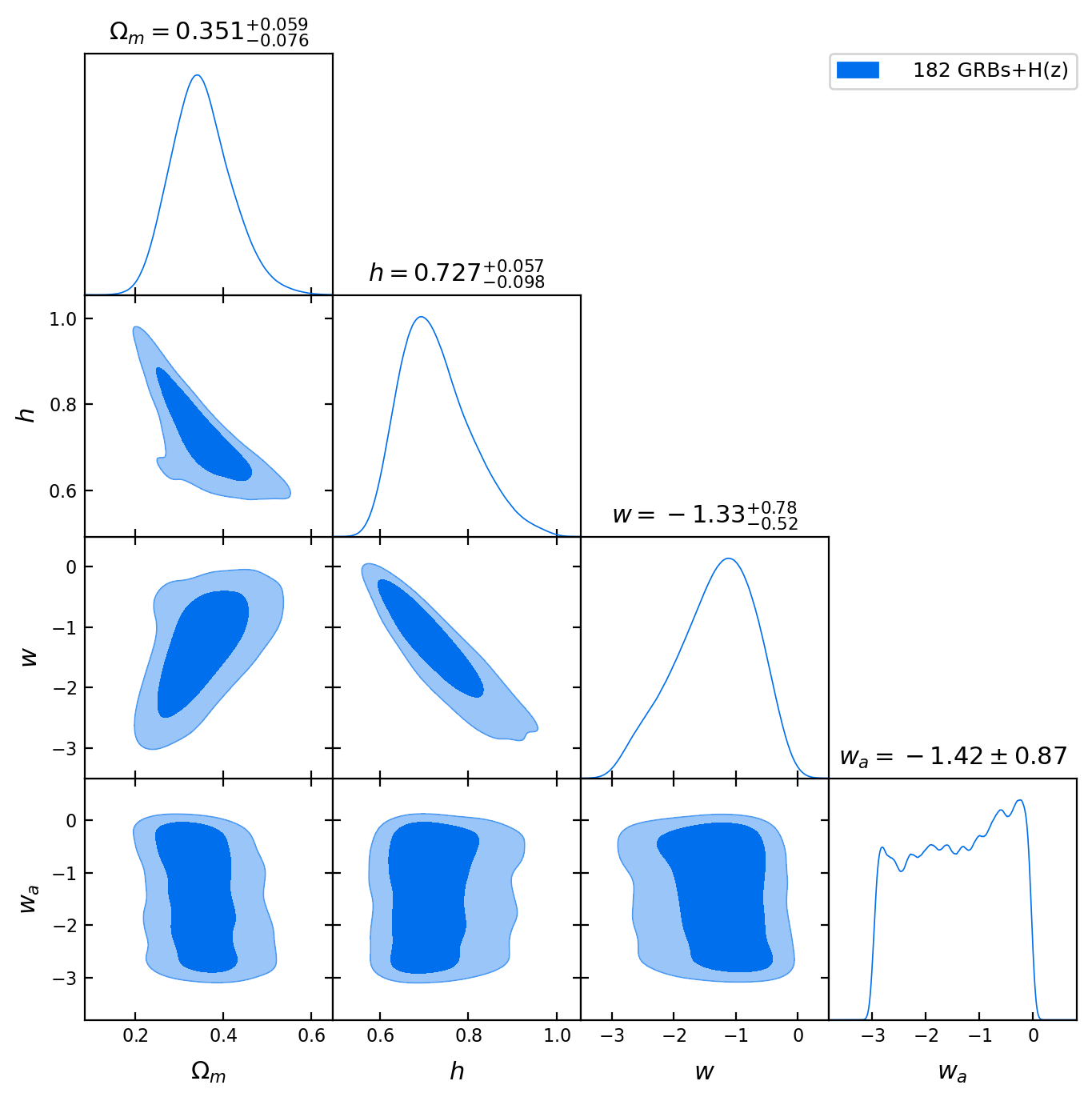}
\includegraphics[width=175px]{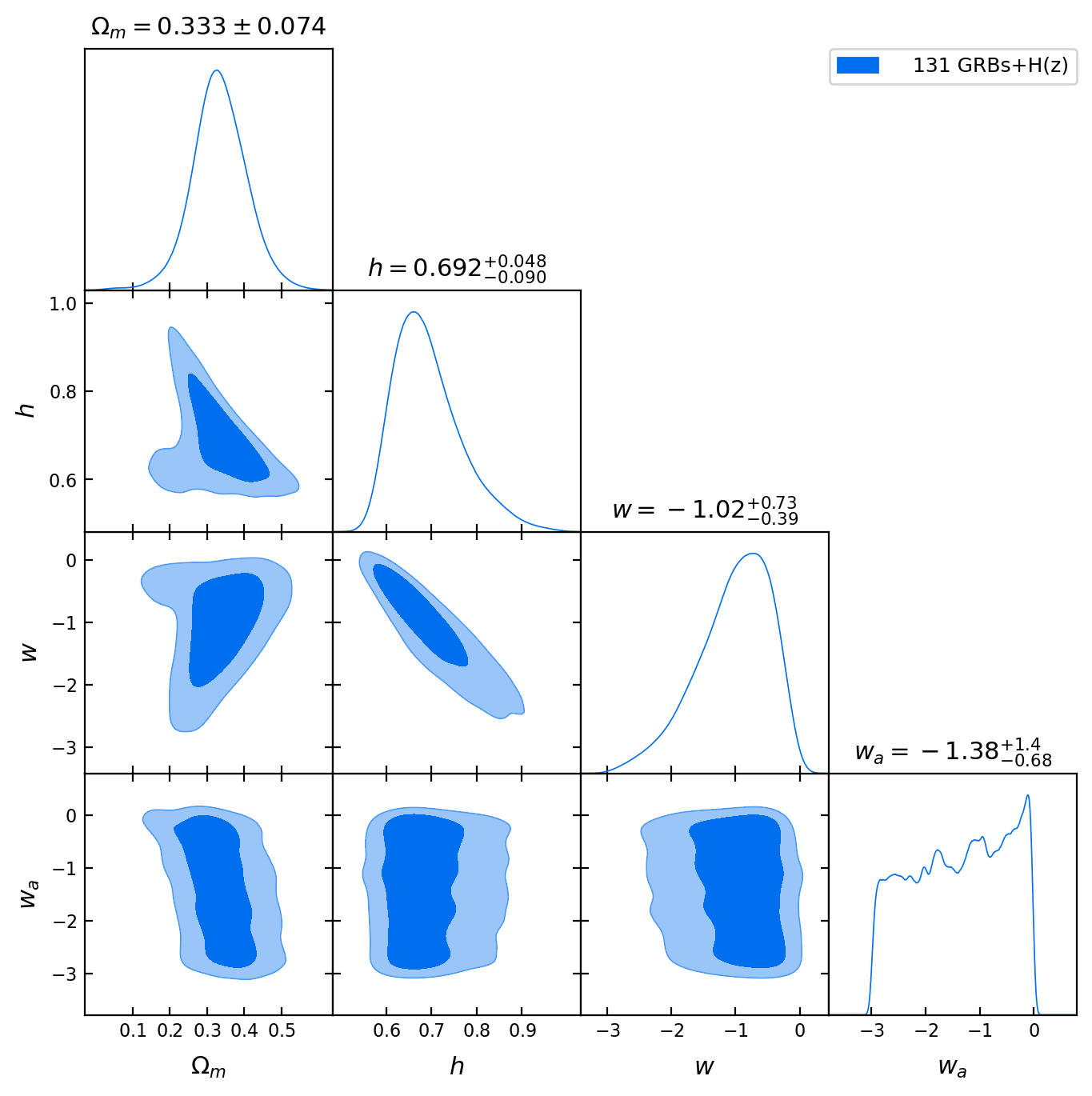}
\caption{Joint constraints on parameters of $\Omega_m$, $h$, $w$ and $w_a$ in the CPL model at redshift $z \geq 0.8$ and $z \geq 1.4$, with 182 GRBs + 32 OHD (left panel) and 131 GRBs + 32 OHD (right panel), respectively from J221 GRBs dataset.}\label{CPL_constrain}
\end{figure*}

\begin{figure*}
\centering
\includegraphics[width=170px]{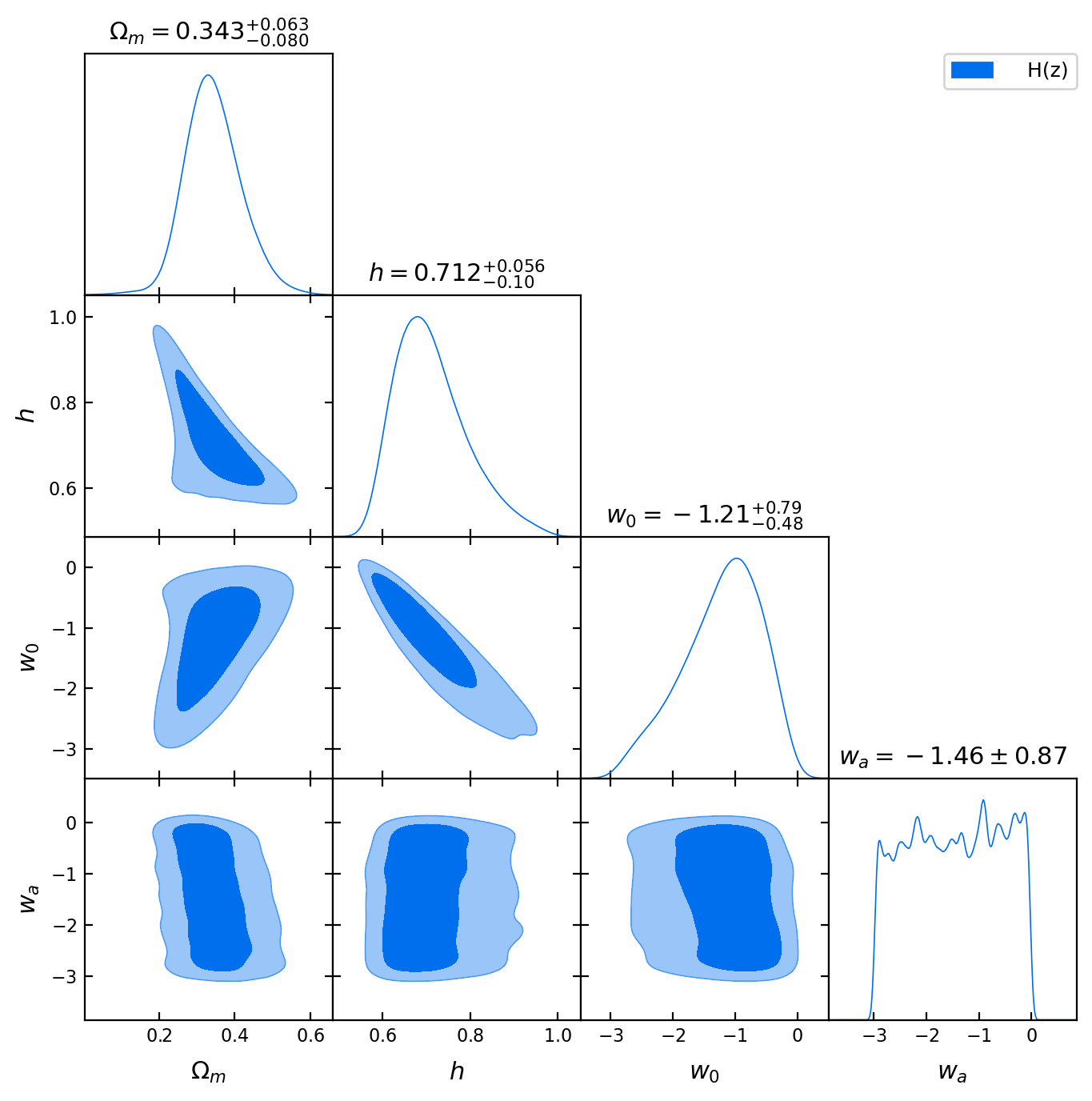}
\caption{Constraints on $\Omega_{\rm m}$, $h$, $w$ and $w_a$ in the CPL models from 32 OHD data set.} \label{constrain_OHD}
\end{figure*}

In order to show the contribution of GRBs to the joint cosmological constraints, we compare the constraint results from OHD with and without GRBs for the CPL model, which are shown in
Fig. \ref{CPL_constrain} and Fig. \ref{constrain_OHD}, 
respectively.
For OHD with GRBs at $z>0.8$ and GRBs at $z>1.4$, we obtain $w_a=-1.42\pm0.87$ and $w_a=-1.38^{+1.40}_{-0.68}$, respectively.  
For OHD alone, we find $w_a=-1.46\pm0.87$, which indicates that DE does not evolve with redshift at the $1 \sigma$ confidence region. 
We can see that the contribution of GRBs to the joint cosmological constraints is a slight shift of adding the best-fit value of
$w_a$ to enclose the $\Lambda$CDM model ($w_a=0$).

\begin{table}
\tbl{Constraints on the $\Lambda$CDM, $w$CDM and CPL models at the 1$\sigma$ confidence level from J221 GRBs at high redshift $z \geq 0.8$(182 GRBs), and $z \geq 1.4$(131 GRBs) with 32 OHD Data Sets. (For the cases only with GRBs, h is set to be 0.7.) }
{\begin{tabular}{@{}cccccccc@{}} \toprule
Data sets & models &$\Omega_{\rm{m}}$& $h$& $w$ & $w_a$ & $\Delta$AIC & $\Delta$BIC\\
\hline \hline
\multirow{3}{*}{182 GRBs ($z\geq0.8$)}& $\Lambda$CDM &  $0.374^{+0.047}_{-0.058}$  & $-$& $-$ & $-$ & $-$ & $-$\\
\multirow{4}{*}{}& $w$CDM & $0.316^{+0.190}_{-0.094}$& $-$&$-1.00^{+0.65}_{-0.28}$& $-$& $2.942$ & $6.147$\\
\multirow{4}{*}{}& CPL& $0.403^{+0.140}_{-0.057}$& $-$&$-1.32^{+1.20}_{-0.98}$& $-1.46^{+1.30}_{-0.71}$& $5.126$ & $11.535$ \\
\hline

\multirow{3}{*}{131 GRBs ($z\geq1.4$)}& $\Lambda$CDM & $0.373^{+0.047}_{-0.058}$  & $-$& $-$& $-$& $-$ & $-$\\
\multirow{4}{*}{}& $w$CDM & $0.21^{+0.14}_{-0.11}$ & $-$ & $-0.98^{+0.54}_{-0.20}$& $-$& $2.687$ & $4.152$\\
\multirow{4}{*}{}& CPL & $0.269^{+0.140}_{-0.076}$ & $-$ & $-1.11^{+1.10}_{-0.39}$& $-1.43^{+1.30}_{-0.59}$& $0.24$ & $5.99$\\
\hline


\multirow{3}{*}{182 GRBs+OHD}&  $\Lambda$CDM & $0.348^{+0.048}_{-0.066}$  & $0.680^{+0.029}_{-0.029}$& $-$& $-$& $-$ & $-$\\
\multirow{4}{*}{}& $w$CDM & $0.318^{+0.067}_{-0.059}$ & $0.704^{+0.055}_{-0.068}$ & $-1.21^{+0.32}_{-0.67}$& $-$& $2.77$ & $6.137$\\
\multirow{4}{*}{}& CPL & $0.351^{+0.059}_{-0.076}$ & $0.727^{+0.057}_{-0.098}$ & $-1.33^{+0.78}_{-0.52}$& $-1.42\pm0.87$& $4.219$ & $10.953$\\
\hline

\multirow{3}{*}{131 GRBs+OHD}&  $\Lambda$CDM & $0.314^{+0.046}_{-0.063}$  & $0.681^{+0.029}_{-0.029}$& $-$& $-$& $-$ & $-$\\
\multirow{4}{*}{}& $w$CDM & $0.269^{+0.100}_{-0.055}$ & $0.683^{+0.042}_{-0.072}$ & $-1.00^{+0.63}_{-0.29}$& $-$& $4.217$ & $7.312$ \\
\multirow{4}{*}{}& CPL& $0.333\pm0.074$ & $0.692^{+0.048}_{-0.090}$ & $-1.02^{+0.73}_{-0.39}$& $-1.38^{+1.40}_{-0.68}$& $4.623$ & $10.811$\\

\botrule
\end{tabular} \label{tab2}}
\end{table}

We also calculate the differences in the Akaike information criterion (AIC; \cite{Akaike1974, Akaike1981}) and the Bayesian information criterion (BIC; \cite{Schwarz1978}) to compare the performance of various models using different combinations of data sets.
\begin{gather}\label{}
	\label{eqnarray_AIC} \mathrm{AIC} = 2p - 2\ln(\mathcal{L}) \\
	\label{eqnarray_BIC} \mathrm{BIC} = p\ln N - 2\ln(\mathcal{L})
\end{gather}
where $p$ is the number of free parameters in a model, $N$ is the sample size of the observational data combination and $\mathcal{L}$ is the maximum value of the likelihood function. The values of $\Delta$AIC and $\Delta$BIC are given by AIC = $\chi^2_{min} + 2\Delta N$, and BIC = $\chi^2_{min} + 2\Delta N$. For the values of AIC and BIC , $0 < \Delta AIC(\Delta BIC) < 2$ indicates  the difficulty in preferring a given model over another, $2 < \Delta AIC(\Delta BIC) < 6$ suggests mild evidence against the model, and $\Delta AIC(\Delta BIC) > 6$ indicates strong evidence against the model. The values of $\Delta$AIC and $\Delta$BIC 
respect to the reference model (the $\Lambda$CDM model) are summarized in Tab. \ref{tab2}. We find that the $\Lambda$CDM model is favored with respect to the $w$CDM model and the CPL model for BIC, which are consistent with the previous analyses \cite{LZL2023}.

\begin{figure*}
\centering
\includegraphics[width=175px]{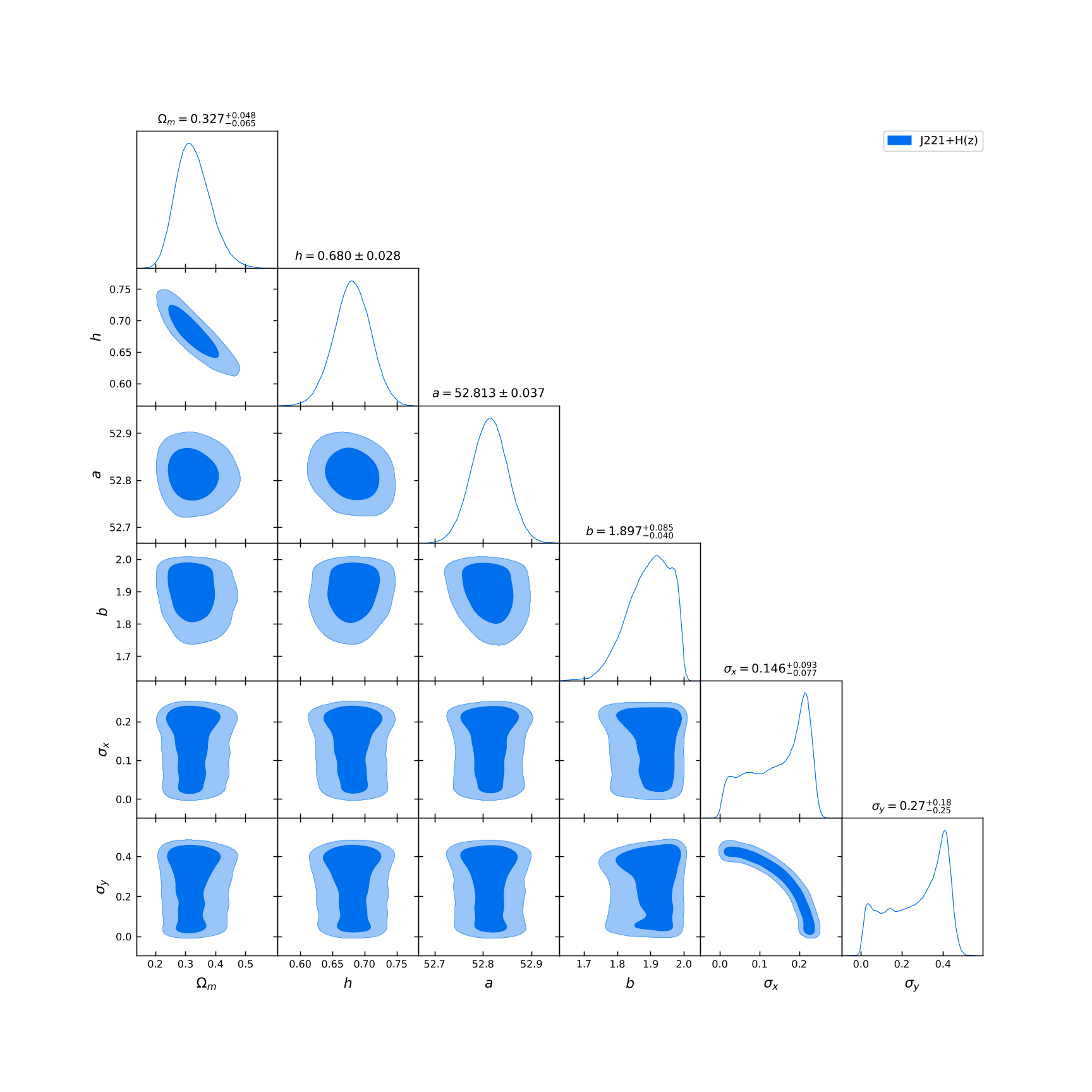}
\includegraphics[width=175px]{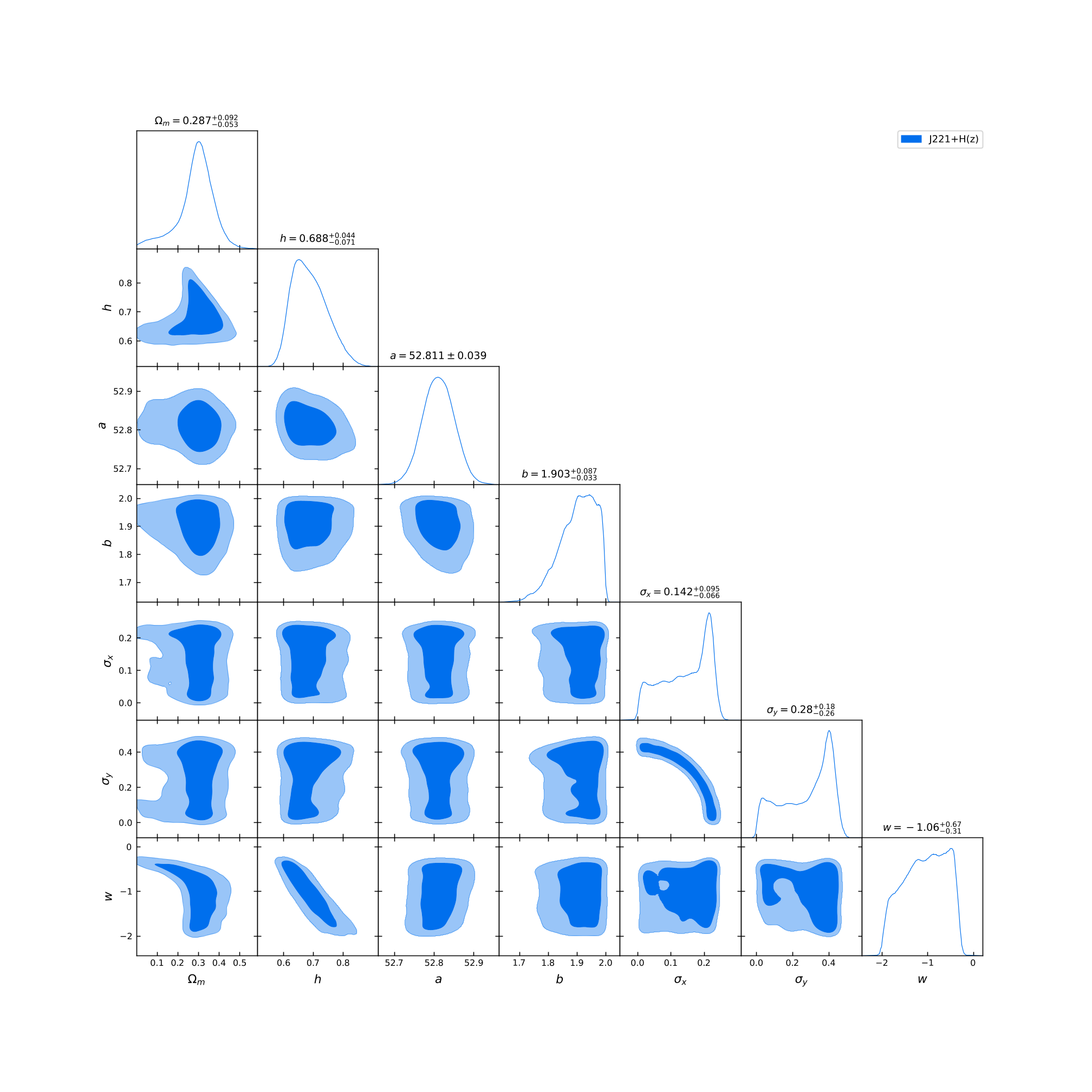}
\caption{Simultaneous fitting parameters of $\Omega_m$, $h$, $a$, $b$, $\sigma_{x}$ and $\sigma_{y}$ in the $\Lambda$CDM model with J221 GRBs + 32 OHD (left panel), and $\Omega_m$, $h$, $a$, $b$, $\sigma_{x}$, $\sigma_{y}$ and $w$ in the $w$CDM model with J221 GRBs + 32 OHD (left panel) (right panel).}\label{constrain_simul}
\end{figure*}

Finally, we also use the J221 data set to constrain the $\Lambda$CDM and $w$CDM models by using the method of simultaneous fitting, in which the parameters of cosmological models ($\Omega_{\rm m}$, $h$, and $w$) and the relation parameters ($a$ and $b$) are fitted simultaneously.
The results from the J221 sample combined with the OHD data set are shown in Fig. \ref{constrain_simul}, and summarized in Tab. \ref{tab3} with the 1$\sigma$ confidence level.
It is found that the values of the coefficients of the Amati relation ($a$, $b$, $\sigma_{{\rm int}}$) for the flat $\Lambda$CDM model and  the flat $w$CDM model in simultaneous fitting are almost identical, which are consistent with the results calibrating from the low-redshift data.
Compared to the results  of Ref.~\refcite{Dainotti2022PASJ74} from GRBs+BAOs with EV ($\Omega_{\rm m}=0.286\pm0.015$, $H_0=67.219\pm1.050\ {\rm km}\ {\rm s}^{-1}{\rm Mpc}^{-1}$), Ref.~\refcite{Dainotti2023MN518} from SNe Ia+BAO+GRBs using GRBs with the correction for the evolution indicated with EV ($\Omega_{\rm m}=0.310\pm0.007$, $H_0=67.83\pm0.16\ {\rm km}\ {\rm s}^{-1}{\rm Mpc}^{-1}$; and $w$ = $-1.017\pm0.015$ for $w$CDM, fixing $\Omega_{\rm m}=0.3$, $h=0.70$), and Ref.~\refcite{Cao2022MN512} from GRBs+BAOs ($\Omega_{\rm m}=0.299^{+0.016}_{-0.018}$, $H_0=69.4\pm1.81\ {\rm km}\ {\rm s}^{-1}{\rm Mpc}^{-1}$),
we find the result of $h$ with J221+OHD for a flat $\Lambda$CDM cosmology is consistent with Ref.~\refcite{Dainotti2022PASJ74,Dainotti2023MN518} and Ref.~\refcite{Cao2022MN512} at the 1$\sigma$ confidence level; and the result of $\Omega_{\rm m}$ is slightly different with Ref.~\refcite{Dainotti2022PASJ74,Dainotti2023MN518} and Ref.~\refcite{Cao2022MN512} at the 1$\sigma$ confidence level. We also find that the value of $w$ for a flat $w$CDM model with  J221+OHD is consistent with Ref.~\refcite{Dainotti2023MN518} at the 1$\sigma$ confidence level.
Following the same approach as in Ref.~\refcite{Dainotti2022MN514,Dainotti2022PASJ74,Dainotti2023ApJ951,Bargiacchi2023MN521}, we also consider the selection biases and redshift evolution for the J221 sample, and find that the fitting results of the cosmological parameters with and without correcting for the evolutionary effects for GRBs are almost identical.


\begin{table}
\tbl{Simultaneous fitting results of $\Omega_{\rm m}$, $h$, $w$, $a$, $b$ and $\sigma_{\rm int}$ in the $\Lambda$CDM and $w$CDM Models, with  J221 GRB + 32 OHD Data Sets.}
{\begin{tabular}{@{}cccccccc@{}} \toprule
Models& Data Sets& $\Omega_{m}$& $h$& $w$& $a$& $b$& $\sigma_{\rm int}$\\ \hline
\multirow{2}{*}{$\Lambda$CDM}&\multirow{2}{*}{J221 GRB + 32 OHD}&\multirow{2}{*}{$0.327^{+0.048}_{-0.065}$}&\multirow{2}{*}{$0.680\pm0.028$}&\multirow{2}{*}{-}&\multirow{2}{*}{$52.813\pm0.037$}&\multirow{2}{*}{$1.897^{+0.085}_{-0.040}$}&\multirow{2}{*}{$0.387$}\\

\multirow{4}{*}{$w$CDM}&\multirow{4}{*}{J221 GRB + 32 OHD}&\multirow{4}{*}{$0.287^{+0.092}_{-0.053}$}&\multirow{4}{*}{$0.688^{+0.044}_{-0.071}$}&\multirow{4}{*}{$-1.06^{+0.67}_{-0.31}$}&\multirow{4}{*}{$52.811\pm0.039$}&\multirow{4}{*}{$1.903^{+0.087}_{-0.033}$}&\multirow{4}{*}{$0.389$}\\
\\
\\
\toprule
\end{tabular} \label{tab3}}
\end{table}

\section{SUMMARY AND DISCUSSION}

In this paper, we use the Gaussian process to calibrate the Amati relation of GRBs 
from the Pantheon+ sample \cite{Scolnic2022} by Gaussian process and obtain the GRB Hubble diagram with the latest J221 GRB sample \cite{Jia2022}. 
With 131 GRBs at $1.4\le z\le8.2$ in the J221 sample and 32 OHD, we obtained $\Omega_{\rm m}$ = $0.314^{+0.046}_{-0.063}$ and $h$ = $0.681^{+0.029}_{-0.029}$ for the flat $\Lambda$CDM model, and  $\Omega_{\rm m}$ = $0.269^{+0.100}_{-0.055}$, $h$ = $0.683^{+0.042}_{-0.072}$, $w$ = $-1.00^{+0.63}_{-0.29}$ for the flat $w$CDM model at the 1$\sigma$ confidence level.
We find that our results with 131 GRBs at $1.4\le z\le8.2$ are consistent with previous analyses obtained in Ref.~\refcite{Liang2022}.
We also use GRB data sets of J221 sample to fit $\Omega_{\rm m}$, $h$, $a$, $b$, $\sigma_{\rm int}$ and $w$ parameters simultaneously. It is found that the simultaneous fitting results are consistent with those obtained from the low-redshift calibration method.

$H_0$ with a redshift evolving is an interesting idea for the $H_0$ tension~\cite{Dainotti2021ApJ912,Dainotti2022Galaxies,Colgain2022,Jia2023,Hu2023,Malekjani2023}.
Recently, Dainotti \emph{et al.} \cite{Dainotti2021ApJ912,Dainotti2022Galaxies} fit the $H_0$ values with a function mimicking the redshift evolution to find a slowly decreasing trend.
Jia \emph{et al.} \cite{Jia2023} find a decreasing trend in the Hubble constant with a significance of a $5.6\sigma$ confidence level with SN Ia, OHD  and baryon acoustic oscillation (BAO) data, which indicates that $H_0$ value is consistent with that measured from the local data at low redshift  and drops to the value measured from the CMB at high redshift.
Malekjani \emph{et al.} \cite{Malekjani2023} find the evolving ($H_0$, $\Omega_{\rm m}$) values above $z = 0.7$ in Pantheon+ sample.
We find that the $H_0$  value with GRBs at $0.8\le z\le8.2$ and OHD at $z\le1.975$  seems to favor the one from the Planck observations, and the $\Omega_{\rm m}$ value of our results for the flat $\Lambda$CDM model is consistent with the CMB observations at the 1$\sigma$ confidence level. A larger $\Omega_{\rm m}$ value in the $\Lambda$CDM model with GRBs at high redshift is obtained, but adding OHD at low redshift  removes this trend.

It should be noted that the combining data sets from different telescopes may introduce an unknown selection bias due to different thresholds and spectroscopy sensitivity \cite{Butler2007,Nava2008}. Examinations of thresholds effects should be required for considering GRBs as joint constraints.
Recent investigations by the DESI collaboration \cite{Colgain2024,Carloni2024,Luongo2024} are particularly interesting, which have shown evidence of potential variations in DE closer examination against the $\Lambda$CDM model.
Future research should be investigated with GRBs at high redshift to provide a more comprehensive understanding.
Furthermore, the potential use of machine learning (ML) algorithms for reconstructing light curves could further enhance parameter determination\cite{Dainotti2023ApJS267}.
Dainotti \emph{et al.} \cite{Dainotti1907} use ML to infer redshifts from the observed temporal and spectral features of GRBs.
Moreover, ML has been used to calibrate the Amati relation \cite{Luongo2021,Zhang2312}.
Dainotti \emph{et al.} \cite{Dainotti2022MN514,Dainotti2022Galaxies} investigate  perspective of the future contribution of GRB-Cosmology.
In the future, GRBs could be used to set tighter constraints on cosmological models 
from Fermi data with much smaller scatters, as well as the data from the Chinese-French mission SVOM (the Space-based multiband astronomical Variable Objects Monitor)\cite{Bernardini2021}, which will provide a substantial enhancement of the number of GRBs  with measured redshift and spectral parameters.

\section*{ACKNOWLEDGMENTS}
We thank the referees for helpful comments and constructive suggestions.
This project was supported by the Guizhou Provincial Science and Technology Foundation (QKHJC-ZK[2021] Key 020 and QKHJC-ZK[2024] general 443).



\end{document}